\documentclass{aa}  

\usepackage{graphicx}
\usepackage{txfonts}
\usepackage{natbib,twoopt}
\usepackage{hyperref} 
\bibpunct{(}{)}{;}{a}{}{,} 
\makeatletter
\newcommandtwoopt{\citeads}[3][][]{\href{http://adsabs.harvard.edu/abs/#3}%
{\def\hyper@linkstart##1##2{}%
\let\hyper@linkend\@empty\citealp[#1][#2]{#3}}}
\newcommandtwoopt{\citepads}[3][][]{\href{http://adsabs.harvard.edu/abs/#3}%
{\def\hyper@linkstart##1##2{}%
\let\hyper@linkend\@empty\citep[#1][#2]{#3}}}
\newcommandtwoopt{\citetads}[3][][]{\href{http://adsabs.harvard.edu/abs/#3}%
{\def\hyper@linkstart##1##2{}%
\let\hyper@linkend\@empty\citet[#1][#2]{#3}}}
\newcommandtwoopt{\citeyearads}[3][][]%
{\href{http://adsabs.harvard.edu/abs/#3}
{\def\hyper@linkstart##1##2{}%
\let\hyper@linkend\@empty\citeyear[#1][#2]{#3}}}
\makeatother

\hypersetup{colorlinks=true,linkcolor=blue,citecolor=blue,urlcolor=blue}

\newcommand{\vra}{\rho}  
\newcommand{\vrs}{v_r}  

\begin{document}

   \title{Astrometric radial velocities for nearby stars}
   \author{Lennart Lindegren \and Dainis Dravins}

   \institute{Lund Observatory, Department of Astronomy and Theoretical Physics, Lund University,
   Box 43, 22100 Lund, Sweden\\
              \email{lennart@astro.lu.se, dainis@astro.lu.se}
             }

   \date{Received XXX YY, 2021; accepted ZZZ WW, 2021}
 
\abstract%
{Under certain conditions, stellar radial velocities can be determined from astrometry, without any use of spectroscopy. This enables us to identify phenomena, other than the Doppler effect, that are displacing spectral lines.}
{The change of stellar proper motions over time (perspective acceleration) is used to determine
radial velocities from accurate astrometric data, which are now available from the \textit{Gaia} and \textsc{Hipparcos} missions.}
{Positions and proper motions at the epoch of \textsc{Hipparcos} are compared with values propagated back from the epoch of the \textit{Gaia} Early Data Release~3. This propagation depends on the radial velocity, which obtains its value from an optimal fit assuming uniform space motion relative to the solar system barycentre.}
{For 930 nearby stars we obtain astrometric radial velocities with formal uncertainties better than 100~km\,s$^{-1}$; for 55 stars the uncertainty is below 10~km\,s$^{-1}$, and for seven it is below 1~km\,s$^{-1}$. Most stars that are not components of double or multiple systems show good agreement with available spectroscopic radial velocities.}
{Astrometry offers geometric methods to determine stellar radial velocity, irrespective of complexities in stellar spectra. This enables us to segregate wavelength displacements caused by the radial motion of the stellar centre-of-mass from those induced by other effects, such as gravitational redshifts in white dwarfs.}


   \titlerunning{Astrometric radial velocities} 
   \authorrunning{L.~Lindegren \& D.~Dravins}

   \maketitle
%

\section{Introduction}
\label{sec:intro}

Stellar radial velocities of enhanced precision are required for applications such as the tracing 
of stellar wobble caused by an exoplanet, motion against a binary companion, or velocity 
relative to nearby stars in a moving cluster. The common method of interpreting wavelength 
positions of spectral lines in terms of a Doppler shift caused by radial motion reaches a limit 
when accuracies much better than $\sim$1~km\,s$^{-1}$ are required.  Even for objects with 
well-defined and rich spectra, limits are set by spectral line asymmetries and displacements 
caused by physical motions on the stellar surface and by gravitational redshifts.  
One step towards an improved understanding of such effects and their eventual mitigation is to 
measure stellar radial velocities also without the use of spectroscopy. With adequate accuracy 
and sufficient baselines in time, such determinations are now enabled through space
astrometry.

Astrometric methods to determine the radial component of stellar motion include monitoring 
the secular change of the annual parallax, measuring changes in proper motion, and assessing 
the varying angular extent of moving clusters whose stars share the same space velocity 
\citepads{1999A&A...348.1040D}. Among these methods, the moving-cluster one offered 
the highest accuracy based on \textsc{Hipparcos} data \citepads{1997A&A...323L..49P},
and radial motions for stars in such clusters have been determined by 
\citetads{2001A&A...367..111D}, \citetads{2019MNRAS.483.5026L}, 
\citetads{2000A&A...356.1119L}, and \citetads{2002A&A...381..446M}. 
When combining astrometric data with spectroscopic data, it then becomes possible to search for 
phenomena, other than the Doppler effect, that are displacing stellar spectral lines 
(\citeads{2003A&A...401.1185L}; \citeads{2003A&A...411..581M}; \citeads{2021arXiv210201079M}; 
\citeads{2011A&A...526A.127P}; \citeads{2002A&A...386..280P}).  
In solar-type stars, both convective blueshifts 
and gravitational redshifts amount to $\sim$0.5 km\,s$^{-1}$, while the moving-cluster method 
enables accuracies of the order of a few hundred m\,s$^{-1}$.  Such levels are comparable 
to the modulation of apparent radial velocities by stellar magnetic activity 
\citepads{2021arXiv210406072M}.

\section{Perspective change of proper motions}
\label{sec:propermotions}

In this paper we examine how perspective acceleration, measured as proper motions changing over time, permits the radial component of stellar motion to be determined. The potential of such astrometric observations was realised already long ago 
(\citeads{1902Vierteljahrsschrift....37...242S}; \citeads{1917AJ.....30..137S}; 
\citeads{1900AN....154...65S}); however, except for a few stars of very large proper motion, they were not practical to apply until adequate observational precision was achieved by space astrometry.  By combining positions and proper motions from \textsc{Hipparcos} with old measurements from the {\textit{Carte du Ciel}} and its Astrographic Catalogue, \citetads{1999A&A...348.1040D} 
obtained astrometric radial velocities for 16 stars with typical errors in the range of 30--40 km\,s$^{-1}$.  For one object, however, the data could be combined with the visual observations by Bessel from 1838, reducing the uncertainty to 11 km\,s$^{-1}$. Recent data releases from the \textit{Gaia} mission have enabled order-of-magnitude improvements over the past \textsc{Hipparcos} study, and they are the topic of this paper.    

Proper motions of stars change gradually with time even if their space motions are strictly uniform relative to the solar system barycentre (SSB). This perspective (or secular) acceleration is proportional to the radial velocity of the star and therefore allows the line-of-sight 
component of the space motion to be measured.  An assumption is that the space motion is not significantly accelerated by the gravitational pull of a companion body. Because the perspective effect is proportional to the parallax and proper motion of the star, the method is limited to 
nearby single stars of high proper motion in practice. Contrary to the moving-cluster method, where the attainable precision is ultimately limited by internal motions in the cluster, the perspective acceleration method continues to improve for measurements accumulated over longer periods in time and, therefore, has a higher potential accuracy, albeit only for nearby, unperturbed stars.

\section{Method and data}
\label{sec:method}

We applied the method to a sample of \textsc{Hipparcos} stars that also appear in the 
\textit{Gaia} Early Data Release~3 (EDR3; \citeads{2021A&A...649A...1G}). The nearly 
25~year epoch difference between \textsc{Hipparcos} (J1991.25) and EDR3 (J2016.0) is 
sufficient to give a potentially interesting precision of the astrometric radial velocity for 
hundreds of stars, although for many of them the measured acceleration is dominated 
by other effects.

Geometric effects cause the proper motion ($\mu$) and parallax ($\varpi$) of nearby 
stars to change at the rates
\begin{equation}\label{eq04}
\dot{\mu}=-2\mu\varpi\vra/A \, , \quad \dot{\varpi}=-\varpi^2\vra/A \, 
\end{equation}  
(e.g.\ \citeads{1917AJ.....30..137S}; \citeads{1977VA.....21..289V}; 
\citeads{1999A&A...348.1040D}),
where $\vra$ is the radial velocity and $A$ is the astronomical unit.%
\footnote{With radial velocity expressed in km\,s$^{-1}$ and the astrometric quantities 
in milliarcseconds (mas), mas per Julian year (yr), and mas\,yr$^{-2}$, we have 
$A=(149597870.7~\text{km})\times(648000000/\pi~\text{mas\,rad}^{-1})
/(365.25\times 86400~\text{s\,yr}^{-1})=9.7779222168\text{\dots}
\times 10^8$~mas\,km\,yr\,s$^{-1}$.}
Following \citetads{2003A&A...401.1185L}, we use $\vra$ to designate the radial component
of the space motion and $\vrs$ for the (spectroscopic) radial velocity.
The perspective acceleration refers to the effect given by $\dot{\mu}$ in Eq.~(\ref{eq04}).
Propagated over the time interval $\Delta t$ ($=24.75$~yr in this case), the accumulated 
change in proper motion is $-2\mu\varpi\vra\Delta t/A$, in position 
$-\mu\varpi\vra\Delta t^2/A$, and in parallax $-\varpi^2\vra\Delta t/A$.
The largest changes are expected for Barnards's star (HIP~87937) owing to its sizeable 
parallax ($\simeq 547$~mas), proper motion ($\simeq 10\,393$~mas\,yr$^{-1}$), and radial velocity 
($\simeq -110$~km\,s$^{-1}$). For this star the perspective effects produce, over the 24.75~yr, 
a position difference of about 393~mas, an increase in the parallax by 0.84~mas, and an increase 
in the proper motion by about 32~mas\,yr$^{-1}$.

Our method to obtain an astrometric estimate of $\vra$ is simple in principle: for a given star, 
the astrometric parameters, as given in EDR3 for the epoch J2016.0, are propagated back in time 
to epoch J1991.25, where they are compared with the corresponding values in the \textsc{Hipparcos} 
catalogue, yielding the differences 
$\Delta\alpha\cos\delta$, $\Delta\delta$, $\Delta\varpi$, $\Delta\mu_{\alpha*}$, and
$\Delta\mu_{\alpha*}$ in the five astrometric parameters. The propagation depends on the assumed value of $\vra$ (at the epoch J2016.0), which is then adjusted for the best overall agreement with the data. We minimised the goodness-of-fit measure
\begin{equation}\label{eq06}
\chi^2(\vra) = \vec{\Delta}^\prime\left(\vec{C}_\text{H}+\vec{C}_\text{G}\right)^{-1}\vec{\Delta}\, ,
\end{equation}  
where $\vec{\Delta}$ is the $5\times 1$ matrix of the parameter differences, 
$\vec{C}_\text{H}$, $\vec{C}_\text{G}$ are the covariance matrices for the corresponding parameters in the \textsc{Hipparcos} catalogue and the (propagated) \textit{Gaia} data, and 
$\vec{\Delta}^\prime$ is the transpose of $\vec{\Delta}$. Confidence intervals of the 
estimated $\vra$ were obtained from the increase in $\chi^2(\vra)$ around the minimum value; in particular, the 68\% confidence interval ($\pm 1\sigma_\vra$) is where
$\smash{\chi^2(\vra) \le \chi^2_\text{min}+1}$ \citep{book:nr3}. For the \textsc{Hipparcos}
data, we used the re-reduction by \citetads{2007ASSL..350.....V}, with covariances computed as described in Appendix~B of \citetads{2014A&A...571A..85M}.

To identify \textsc{Hipparcos} stars in \textit{Gaia} EDR3, we used the cross-match table
\texttt{gaiaedr3.hipparcos2\_best\_neighbour} provided with the release; of the 99\,525 \textsc{Hipparcos} entries in that table, 98\,004 have valid positions and proper motions in both catalogues. However, it is not meaningful to attempt to estimate the radial velocity for all of them. As indicated by Eq.~(\ref{eq04}), the precision on $\vra$ mainly 
depends on the size of the product $\mu\varpi$, and we therefore consider only the 13\,161 sources in this sample that have $\mu\varpi>1000$~mas$^2$\,yr$^{-1}$ in EDR3. 

The practical implementation of the method to estimate $\vra$ is complicated by the need to use a very accurate algorithm for propagating the astrometric parameters and by possible systematic differences 
between the \textit{Gaia} and \textsc{Hipparcos} data sets. We used the propagation formulae derived by \citetads{2014A&A...570A..62B}, which assume uniform rectilinear motion relative to the SSB and rigorously take light-time effects  into account in addition to the geometric effects in Eq.~(\ref{eq04}).

In \textit{Gaia} EDR3, there are known issues with the parallax zero point as well as with the reference 
frame of the bright stars ($G\lesssim 13$, which includes all \textsc{Hipparcos} stars). Moreover, 
the \textsc{Hipparcos} reference frame may differ systematically from the EDR3 frame even after
the latter has been corrected for the known issues (e.g.\ \citeads{2021arXiv210511662B}). All of this 
could produce errors of the order of 1~mas and 0.2~mas\,yr$^{-1}$ in the calculated parameter 
differences $\vec{\Delta}$ in Eq.~(\ref{eq06}), which would affect our estimation of $\vra$. In order 
to minimise their impact, we used the following correction procedure in three steps. 
In the first step, the EDR3 data were corrected by subtracting the parallax bias according to 
\citetads{2021A&A...649A...4L} and the magnitude-dependent proper motion bias according to 
\citetads{2021A&A...649A.124C}. The resulting EDR3 data are, to the best of our knowledge, absolute
and on a non-rotating reference frame for all magnitudes. 

In the second step, we propagated the corrected EDR3 data to the \textsc{Hipparcos} epoch and
analysed the \textsc{Hipparcos}--\textit{Gaia} differences in position and proper motion in terms
of an orientation error ($\vec{\varepsilon}$) and spin ($\vec{\omega}$) of the \textsc{Hipparcos} 
reference frame with respect to the (corrected) \textit{Gaia} frame. This used the equations
\begin{equation}
\begin{bmatrix} (\alpha_\text{H}-\alpha_\text{G})\cos\delta \\[6pt] 
\delta_\text{H}-\delta_\text{G} \end{bmatrix} = \vec{M}\vec{\varepsilon}\, , \quad\quad 
\begin{bmatrix} \mu_{\alpha*,\text{H}}-\mu_{\alpha*,\text{G}} \\[6pt]
\mu_{\delta,\text{H}}-\mu_{\delta,\text{G}} \end{bmatrix} = \vec{M}\vec{\omega}\, ,
\end{equation}
where the subscript H refers to the \textsc{Hipparcos} data as published by \citetads{2007ASSL..350.....V},
G refers to the corrected and propagated \textit{Gaia} data, and
\begin{equation}
\vec{M} = \begin{bmatrix} 
-\sin\delta\cos\alpha & -\sin\delta\sin\alpha &\cos\delta\\
\phantom{+}\sin\alpha & \cos\alpha & \end{bmatrix}\, .
\end{equation}
Using least-squares solutions weighted by the inverse combined variances, as in
Eq.~(\ref{eq06}), this resulted in the estimates
\begin{equation}\label{eq:eps}
\vec{\varepsilon}(1991.25)=\begin{bmatrix} -0.574\pm 0.007\\
-0.888\pm 0.007\\ +0.243\pm 0.008 \end{bmatrix}~\text{mas}
\end{equation}
and
\begin{equation}\label{eq:ome}
\vec{\omega}=\begin{bmatrix} -0.094\pm 0.004\\ 
+0.198\pm 0.004\\ +0.080\pm 0.005 \end{bmatrix}~\text{mas\,yr$^{-1}$} \, .
\end{equation}
Initially the full set of 98\,004 sources was included; however, after iteratively removing sources 
with significant residuals (mainly binaries), the final solutions used 86\,914 sources for
$\vec{\varepsilon}$ and 96\,400 sources for $\vec{\omega}$. The uncertainties were 
estimated by bootstrap resampling. The offsets in Eqs.~(\ref{eq:eps}) and (\ref{eq:ome}) 
are consistent with the estimated uncertainties of the \textsc{Hipparcos} reference frame,
namely $\pm0.6$~mas for the components of $\vec{\varepsilon}(1991.25)$ and 
$\pm0.25$~mas\,yr$^{-1}$ for the components of $\vec{\omega}$ \citepads{1997A&A...323..620K}.

In the third and final step, the \textsc{Hipparcos} data were transformed to the (corrected) 
\textit{Gaia} reference frame by subtracting $\vec{M}\vec{\varepsilon}$ from the positions
and $\vec{M}\vec{\omega}$ from the proper motions. These were then used together with
corrected \textit{Gaia} data to compute the parameter differences $\vec{\Delta}$ in Eq.~(\ref{eq06}).

The various systematic corrections to the parallaxes and, in particular, the proper motions 
have a small but not entirely negligible impact on the estimated values of $\vra$. However, if 
none of the above corrections were applied, the resulting changes in $\vra$ would in most 
cases be less than $\pm 0.3\sigma_{\vra}$.

We note that other researchers have used similar techniques of combining \textsc{Hipparcos} 
and \textit{Gaia} data for the somewhat orthogonal purpose of identifying astrometric binaries and stars dynamically accelerated by faint companions. Investigations based on the second release (DR2) of \textit{Gaia} data include \citetads{2018ApJS..239...31B} and \citetads{2019A&A...623A..72K}, while 
\citetads{2021arXiv210511662B} provides an update using EDR3. In these studies the perspective
effect was removed by adopting the spectroscopic radial velocity, when available, for the radial motion.
In principle, the resulting $\chi^2$ (or similar) then includes a contribution from a possible 
difference between the astrometric and spectroscopic velocities; however, in general our method 
tends to give a high $\chi^2_\text{min}$ for the same stars that are identified as accelerated in 
these other studies.

\begin{figure}
\centering
  \includegraphics[width=\hsize]{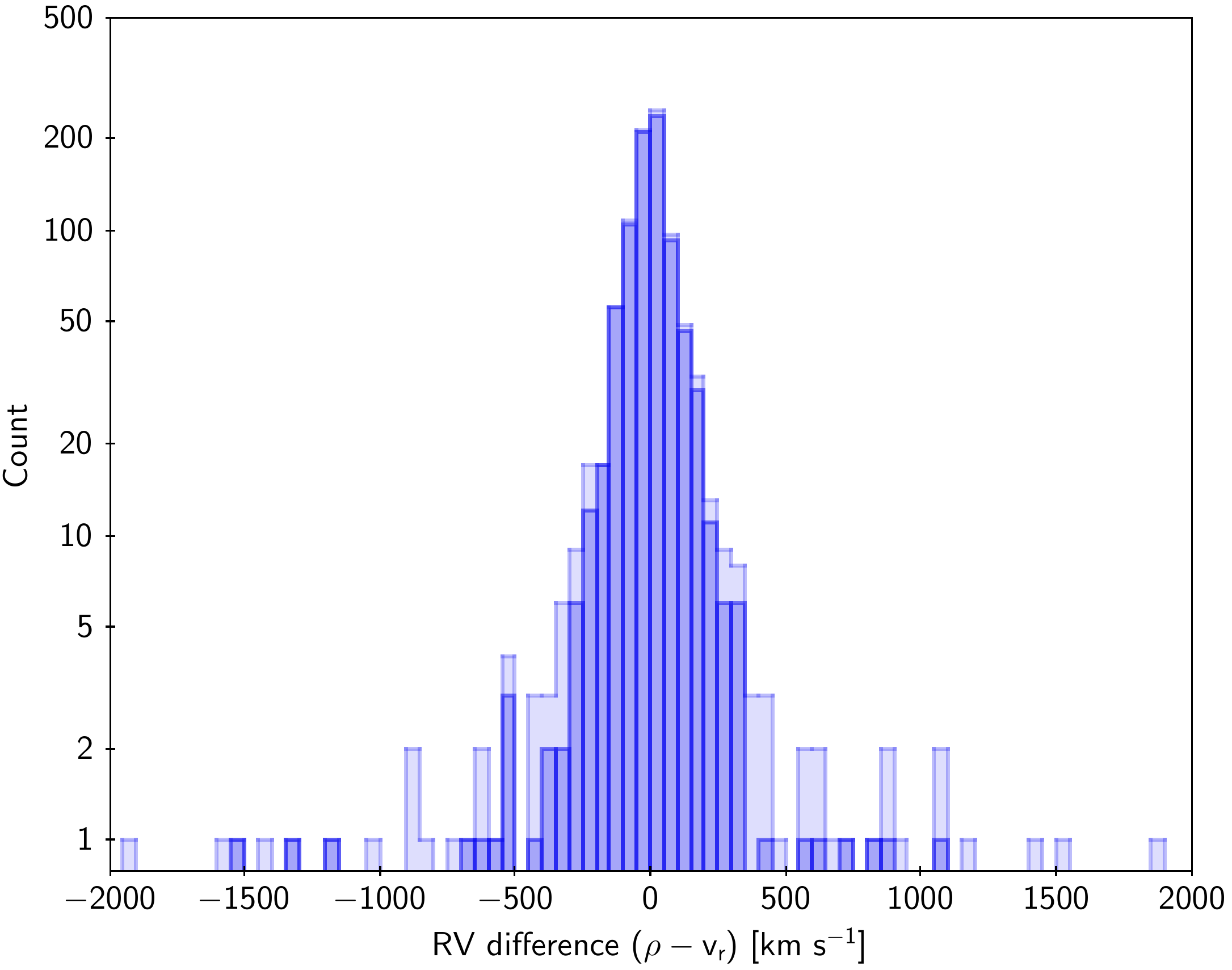}
    \caption{Distribution of the difference $\vra-\vrs$ between the astrometric and 
    spectroscopic radial velocities for two samples of \textsc{Hipparcos} stars. 
    \textit{Light blue:} 926 stars with $|\,\vra\,|<2000$~km\,s$^{-1}$, $\sigma_{\vra}<100$~km\,s$^{-1}$, 
    and with radial velocities in SIMBAD. \textit{Dark blue:} Sub-sample of 852 stars that, in 
    addition, have  $\chi^2_\text{min}<40$. The bin size is 50~km\,s$^{-1}$.}
    \label{fig:hist_DeltaVr}
\end{figure}

\begin{figure}
\centering
  \includegraphics[width=\hsize]{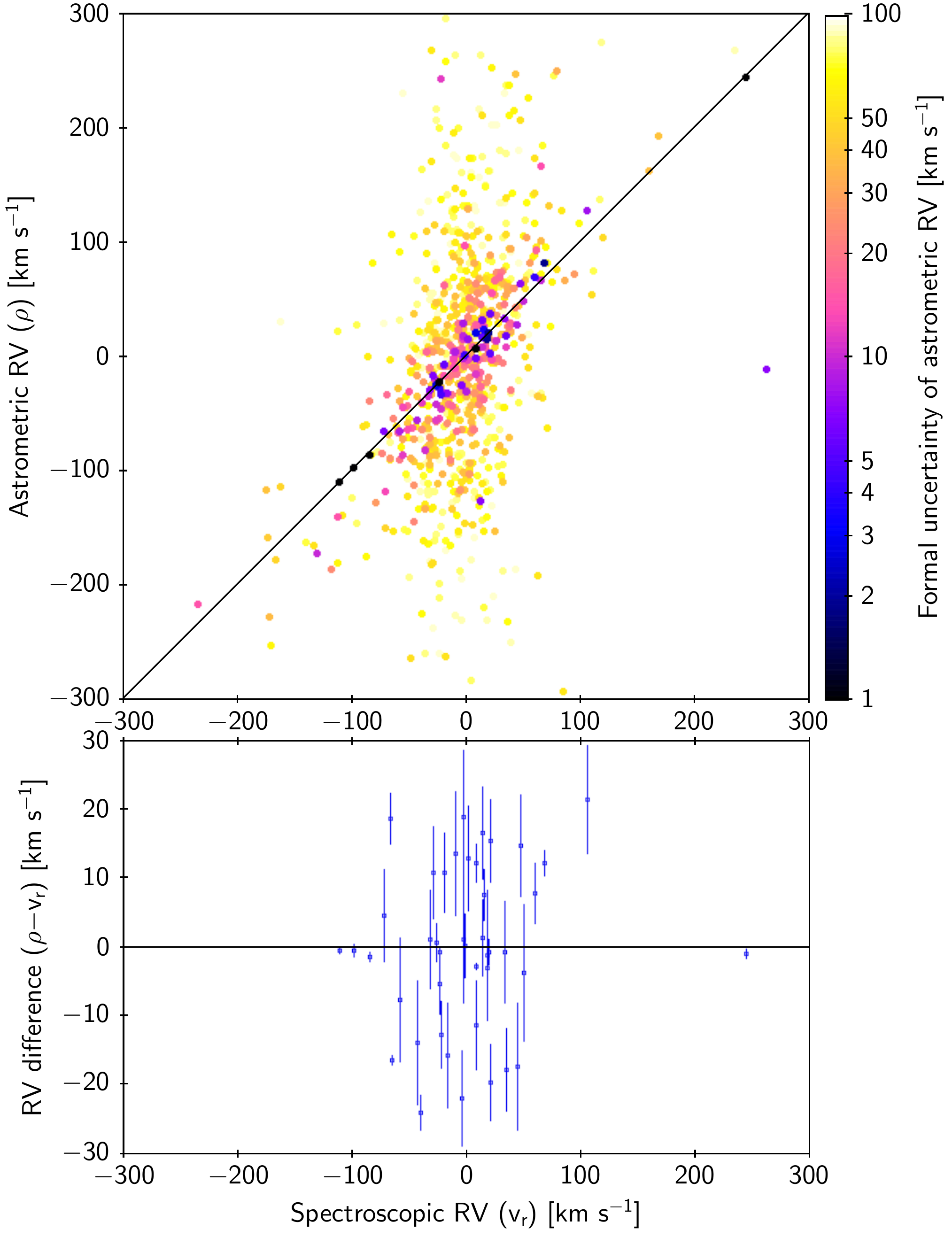}
    \caption{Comparison of the astrometric and spectroscopic radial velocities. \textit{Top:} 
    Astrometric versus spectroscopic values for the sample shown as the dark blue 
    histogram in Fig.~\ref{fig:hist_DeltaVr} and colour-coded according to the formal uncertainty 
    ($\sigma_{\vra}$) of the astrometric radial velocity estimate. The black line is the 1:1 relation.
There are 28 data points that fall outside the limits of this plot. 
    \textit{Bottom:} Difference between the astrometric and spectroscopic velocities for the sub-sample
    with $\sigma_{\vra}<10$~km\,s$^{-1}$ (corresponding to Table~\ref{table1}), with error bars at
    $\pm\sigma_{\vra}$. There are 11 data points outside of this plot.}
    \label{fig:vr_vr_vrE}
\end{figure}

\begin{table*}
\caption{Estimated astrometric radial velocities for selected \textsc{Hipparcos} stars in order of 
increasing formal uncertainty.
\label{table1}}
\small
\begin{tabular}{rllrrrl}
\hline\hline
\noalign{\smallskip}
HIP & Name & Sp 
&$\vrs$\phantom{}
&\multicolumn{1}{c}{$\vra$} 
& $\chi^2_\text{min}$ 
& Remark\\
\noalign{\smallskip}
\hline
\noalign{\smallskip}
87937 &  Barnard's star & M4V & $-110.35$ & $ -110.92\pm 0.37$ &      7.3 & \\
114046 & HD 217987 & M2V & $  +8.82$ & $   +5.86\pm 0.50$ &      9.2 & \\
54035 & HD  95735 & M2+V & $ -84.69$ & $  -86.15\pm 0.68$ &     22.3 & \\
24186 & HD  33793 & M1VIp & $+245.23$ & $ +244.11\pm 0.70$ &      7.4 & Kapteyn's star\\
104217 &   61 Cyg B & K7V & $ -64.25$ & $  -80.86\pm 0.74$ &  29284.9 & see Table~\ref{table2}\\
70890 &  Proxima Centauri & M5.5Ve & $ -22.40$ & $  -23.20\pm 0.74$ &      7.6 & \\
57939 & HD 103095 & K1V\_Fe-1.5 & $ -98.01$ & $  -98.61\pm 0.99$ &     39.3 & Groombridge 1830\\
1475 & V GX And & M2V & $ +11.82$ & $  -57.57\pm 1.28$ &    391.7 & \\
54211 & BD+44  2051 & M1.0Ve & $ +68.75$ & $  +80.97\pm 1.85$ &     17.2 & \\
36208 & BD+05  1668 & M3.5V & $ +18.22$ & $  +15.08\pm 1.87$ &      5.1 & Luyten's star\\
105090 & V AX Mic & M1V & $ +20.56$ & $  +19.82\pm 1.91$ &     12.4 & \\
108870 & $\varepsilon$ Ind & K5V & $ -40.03$ & $  -64.19\pm 2.55$ &    464.1 & \\
49908 & HD  88230 & K6VeFe-1 & $ -25.92$ & $  -25.45\pm 2.75$ &     16.5 & \\
25878 & HD  36395 & M1.5Ve & $  +8.63$ & $  +20.63\pm 2.87$ &      6.8 & \\
67155 & HD 119850 & M2V & $ +15.81$ & $  +23.35\pm 3.61$ &      6.2 & \\
104214 &   61 Cyg A & K5V & $ -65.82$ & $  -47.32\pm 3.69$ &    248.9 & see Table~\ref{table2}\\
55360 & V SZ UMa & M2V & $ +60.43$ & $  +68.24\pm 4.28$ &      4.5 & \\
85295 & HD 157881 & K7V & $ -23.52$ & $  -28.96\pm 4.50$ &      3.0 & \\
84478 & V V2215 Oph & K5V & $  -0.09$ & $   +0.15\pm 4.57$ &     31.6 & \\
80824 & BD$-$12  4523 & M3V & $ -21.22$ & $  -34.11\pm 4.83$ &      0.5 & \\
57367 & LAWD 37 & DQ & -- & $  +28.05\pm 4.91$ &      1.1 & \\
91768 & HD 173739 & M3V & $  -0.77$ & $ +138.52\pm 5.28$ &    564.3 & see Table~\ref{table2}\\
76074 & CD$-$40  9712 & M2.5V & $ +21.46$ & $   +1.70\pm 5.51$ &      6.2 & \\
41926 & HD  72673 & K1V & $ +14.72$ & $  +15.98\pm 5.64$ &     10.2 & \\
114622 & HD 219134 & K3V & $ -18.48$ & $   -7.77\pm 5.79$ &     20.7 & \\
94761 & HD 180617 & M3-V & $ +35.88$ & $  +18.07\pm 5.95$ &      3.2 & \\
23311 & HD  32147 & K3+V & $ +21.62$ & $  +37.02\pm 6.09$ &     15.2 & \\
79537 & HD 145417 & K3VFe-1.7 & $  +8.73$ & $   -2.52\pm 6.46$ &      8.2 & \\
65859 & BD+11  2576 & M1.0Ve & $ +14.56$ & $  +31.01\pm 6.59$ &      2.2 & \\
117473 & V BR Psc & dM1 & $ -71.13$ & $  -66.57\pm 6.65$ &      6.4 & \\
86162 & BD+68   946 & M3.0V & $ -28.76$ & $  -17.96\pm 6.73$ &      3.1 & \\
85523 & CD$-$46 11540 & M3V & $  -2.90$ & $  -25.05\pm 6.85$ &      1.1 & \\
3829 & Wolf   28 & DZ7.5 & $+263$ & $  -11.74\pm 7.11$ &      1.7 & van Maanen 2\\
106440 & HD 204961 & M2/3V & $ +13.16$ & $ -127.25\pm 7.16$ &     33.8 & \\
57548 & Ross  128 & dM4 & $ -31.07$ & $  -30.13\pm 7.18$ &      3.0 & \\
73184 & HD 131977 & K4V & $ +26.99$ & $ +207.16\pm 7.34$ &    120.4 & \\
83591 & HD 154363 & K4/5V & $ +34.03$ & $  +33.34\pm 7.39$ &      8.7 & \\
58345 & HD 103932 & K4+V & $ +48.57$ & $  +63.16\pm 7.41$ &      3.8 & \\
8102 & $\tau$ Cet & G8V & $ -16.60$ & $  -32.31\pm 7.60$ &      5.4 & \\
4856 & Ross  318 & M3.0Ve & $  +1.52$ & $  +14.19\pm 7.65$ &      2.6 & \\
26857 & Ross   47 & M4V & $+105.88$ & $ +127.17\pm 7.92$ &      2.2 & \\
91772 & HD 173740 & M3.5V & $  +1.10$ & $ -209.03\pm 8.26$ &    956.5 & see Table~\ref{table2}\\
98792 & HD 190404 & K1V & $  -2.47$ & $  +28.73\pm 8.74$ &      7.5 & \\
104432 & Wolf  918 & M1V & $ -58.26$ & $  -66.01\pm 9.03$ &      2.9 & \\
86287 & BD+18  3421 & M1.5Ve & $  -9.51$ & $   +4.05\pm 9.03$ &      0.6 & \\
86990 & L  205$-$128 & M3.5V & $ -42.94$ & $  -56.77\pm 9.05$ &      3.1 & \\
74235 & HD 134439 & sd:K1Fe-1 & $+309.99$ & $ +318.52\pm 9.11$ &      2.3 & \\
82588 & HD 152391 & G8.5Vk: & $ +45.12$ & $  +27.75\pm 9.12$ &      9.4 & \\
21088 & G 175$-$34 & M4.0Ve & $ +28.80$ & $  +74.92\pm 9.18$ &    356.6 & \\
112460 & V EV Lac & M4.0V & $  +0.29$ & $  -31.82\pm 9.18$ &      3.4 & \\
10279 & BD+02   348 & M1V & $  -2.56$ & $   -1.46\pm 9.23$ &      1.7 & \\
67090 & BD+18  2776 & M1V & $ +18.90$ & $  +17.61\pm 9.42$ &     18.8 & \\
29295 & HD  42581 & M1V & $  +4.73$ & $-1011.33\pm 9.50$ &   4885.0 & \\
113020 & BD$-$15  6290 & M3.5V & $  -1.60$ & $  +17.33\pm 9.59$ &      3.8 & \\
60559 & Ross  695 & dM2.0 & $ +51.19$ & $  +47.33\pm 9.97$ &      0.6 & \\
\noalign{\smallskip}
\hline
\end{tabular}
\tablefoot{Column~1: \textsc{Hipparcos} identifier.
Columns~2--4: main identifier, spectral type, and (spectroscopic) radial velocity ($\vrs$ in 
km\,s$^{-1}$) as given in SIMBAD.
Column~5: estimated astrometric radial velocity ($\vra$) at epoch J2016.0 and its formal uncertainty 
($\pm\sigma_{\vra}$), both in km\,s$^{-1}$.
Column~6: goodness-of-fit in Eq.~(\ref{eq06}) for the estimated $\vra$ 
(4~degrees of freedom).  
}
\end{table*}

\begin{table*}
\caption{Estimated systemic astrometric radial velocities and mass ratios for physical pairs.
\label{table2}}
\small
\begin{tabular}{lrrrrlr}
\hline\hline
\noalign{\smallskip}
System & HIP (A) & HIP (B) & $\vrs$ (CM) & \multicolumn{1}{c}{$\vra$ (CM)} 
& \multicolumn{1}{c}{$q$}  & \multicolumn{1}{c}{$\chi^2_\text{min}$ (dof)} \\
\noalign{\smallskip}
\hline
\noalign{\smallskip}
GJ 338 AB & 45343 & 120005 & $  +11.51$ & $ +152.82\pm 32.00$ & $0.452\pm 0.092$ &   9.60 (3)\\
 &  &  & $  +11.67$ & $  +50.83\pm 20.76$ & $0.959$ &  26.41 (4)\\
\\
GJ 725 AB & 91768 & 91772 & $   -0.11$ & $  +16.71\pm  6.31$ & $0.544\pm 0.030$ &  49.90 (3)\\
 &  &  & $   +0.04$ & $  -11.27\pm  4.79$ & $0.766$ &  89.71 (4)\\
\\
61 Cyg AB & 104214 & 104217 & $  -65.14$ & $  -65.53\pm  2.25$ & $0.758\pm 0.049$ &   5.39 (3)\\
 &  &  & $  -65.09$ & $  -67.03\pm  2.03$ & $0.877$ &  11.29 (4)\\
\noalign{\smallskip}
\hline
\end{tabular}
\tablefoot{Column~1: name of the system.
Columns~2--3: \textsc{Hipparcos} identifiers of the components.
Column~4: spectroscopic radial velocity (in km\,s$^{-1}$) for the centre of mass (CM), using 
component values from SIMBAD and the given mass ratio.
Column~5: estimated astrometric radial velocity of the centre of mass and its formal uncertainty 
($\pm\sigma_{\vra}$), both in km\,s$^{-1}$.
Column~6: estimated or assumed mass ratio $q=M_\text{B}/M_\text{A}$.
Column~7: goodness-of-fit with the number of degrees of freedom in brackets. For each system, the first 
line shows the result when both $\vra$ and $q$ are fitted, and the second shows the result when $q$ is 
constrained to the value in the line (see text for references).
}
\end{table*}

\section{Determined radial velocities}
\label{sec:results}

When the algorithm outlined above is applied to the selected sample of \textsc{Hipparcos} stars, 
most of them obtain physically unrealistic values of $\vra$ and/or very large formal uncertainties 
$\sigma_{\vra}$. For 930~stars, we find $|\,\vra\,|<2000$~km\,s$^{-1}$ and $\sigma_{\vra}<100$~km\,s$^{-1}$
and in Table~\ref{table1} we give the results for the 55 entries among them
having $\sigma_{\vra}<10$~km\,s$^{-1}$. Of the 930 stars, all but four have spectroscopic radial 
velocities ($\vrs$) in the SIMBAD database \citepads{2000A&AS..143....9W}, and in 
Fig.~\ref{fig:hist_DeltaVr} we show the distribution of 
$\vra-\vrs$ for these stars as the light-blue histogram. Their median $\vra-\vrs$ is 
$+2.3$~km\,s$^{-1}$ and the interquartile range (IQR) is $101.3$~km\,s$^{-1}$.
The sub-sample of 852 stars that also have a reasonably good fit is shown in darker blue, 
$\chi^2_\text{min}<40$, indicating that the stars are perhaps not strongly perturbed by a companion. 
For these stars the median difference is $+1.2$~km\,s$^{-1}$ and the IQR is $95.5$~km\,s$^{-1}$.

For a number of the high-precision stars with small or moderate $\chi^2_\text{min}$, the agreement 
between the astrometric and spectroscopic values is remarkable. This is illustrated in the top
panel of 
Fig.~\ref{fig:vr_vr_vrE}, where the two quantities are plotted against each other for the sub-sample 
with $\chi^2_\text{min}<40$. The correlation is especially clear for the high-precision points 
shown in darker colours. 
Excluding the outlier van~Maanen~2 (see Sect.~\ref{ssec:vanmaanen2}), a 
least-squares fit to the remaining 851 data points, weighted by $\sigma_{\vra}^{-2}$, yields
\begin{equation}\label{eq07}
\vra = (-1.11\pm 0.65~\text{km\,s$^{-1}$}) + (0.9999\pm 0.0060)\,\vrs \, .
\end{equation}
The reduced chi-square of this fit is high, about 8.3, as can be expected from the presence of
a number of binaries in the sample. The closeness of Eq.~(\ref{eq07}) to the equality 
relation $\vra=\vrs$, nevertheless, suggests that the binaries do not bias the population mean of
$\vra$. The lower panel in Fig.~\ref{fig:vr_vr_vrE} shows the velocity differences, with error bars, 
for the sub-sample in Table~\ref{table1}.

The uncertainties of $\vra$ given in the table and diagrams are formal ones obtained as
described in Sect.~\ref{sec:method}, based on the astrometric uncertainties provided in the
\textsc{Hipparcos} and \textit{Gaia} EDR3 catalogues. From a comparison of these two
catalogues, \citetads{2021arXiv210511662B} conclude that the uncertainties in EDR3,
for these bright sources, are underestimated by a factor of 1.37. 
If this is also the case for the
present high-precision sub-sample, the uncertainties in Table~\ref{table1} on
average should be multiplied by a factor $\simeq 1.2$, and $\chi^2_\text{min}$ by a factor 
$\simeq 0.9$. It would however have a negligible impact on the estimated values of $\vra$.
 
Many entries in Table~\ref{table1} show a large discrepancy between $\vra$ and $\vrs$ (by many times 
the formal uncertainty) combined with a large $\chi^2_\text{min}$. These may be components of physical 
systems where the mutual perturbations cause dynamical accelerations that strongly bias the estimated 
$\vra$. A prime example is the binary system 61~Cyg~AB (HIP~104214+104127), where the application of the 
method in Sect.~\ref{sec:method} to the individual components gives $\chi^2_\text{min}$ of 249 and 
29\,285, respectively. 

For binaries such as this, the method should instead be applied to the centre of mass (CM) of the 
system, yielding an estimate of the systemic $\vra$.  Given the mass ratio $q=M_\text{B}/M_\text{A}$, 
the astrometric parameters ($\vec{a}$) of the CM in each catalogue can be computed as 
$(\vec{a}_1+q\vec{a}_2)/(1+q)$, with covariance $(\vec{C}_1+q^2\vec{C}_2)/(1+q)^2$. Propagating these 
parameters from EDR3 to the \textsc{Hipparcos} epoch and applying Eq.~(\ref{eq06}) yields a function 
$\chi^2(\vra,q)$ from which it may be possible to estimate both the (systemic) $\vra$ and the mass 
ratio $q$. 

Because the mass ratio equals the inverse ratio of the dynamical accelerations (e.g.\ Eq.~(24) in \citeads{2019A&A...623A..72K}), it is only possible to determine the mass ratio when the measured accelerations are significantly different for the two components, leading to significantly different (and biased) estimates of $\vra$ when the dynamical contributions are neglected. In addition to 61~Cyg~AB (with $\Delta\vra\equiv\vra_\text{B}-\vra_\text{A} = -33.54\pm 3.76$~km\,s$^{-1}$), there are two other systems that fulfil this condition: HIP~91768+91772 ($\Delta\vra = -347.55\pm 9.80$~km\,s$^{-1}$) and HIP~45343+120005 ($\Delta\vra = -552.68\pm 41.26$~km\,s$^{-1}$).
Estimates of $q$ and the centre-of-mass $\vra$ for these systems are given in Table~\ref{table2}. For comparison, the table also gives the estimated $\vra$ and $\chi^2_\text{min}$ when the mass ratio is constrained to the value based on evolutionary models (for 61~Cyg, from  \citeads{2008A&A...488..667K}), or the absolute $K$ band magnitude (for GJ~338 and GJ725, from \citeads{2019A&A...623A..72K} using the calibration by \citeads{2015ApJ...804...64M}). The $q$ values used in the constrained solutions are given without error limits in the second line of each object in Table~\ref{table2}. For a separate discussion of 61~Cyg~AB, see Sect.~\ref{ssec:61cyg}.

For many more pairs, it is possible to compute a systemic $\vra$ by assuming a particular value of $q$ for the pair, for example,
based on infrared photometry. While these estimates are less biased by dynamical acceleration than the component values $\vra_\text{A}$ and $\vra_\text{B}$ and, therefore, often in much better agreement with the spectroscopic values, we do not report them here as they are to a good approximation given by the weighted averages 
$\vra_\text{CM}=(\vra_\text{A}+q\vra_\text{B})/(1+q)$.

\section{Notes on individual objects}
\subsection{HIP~3829 (van Maanen~2)}
\label{ssec:vanmaanen2}

In Fig.~\ref{fig:vr_vr_vrE}, one strikingly deviating point at $\vrs=263$~km\,s$^{-1}$ and 
$\vra=-12\pm 7$~km\,s$^{-1}$ is van Maanen~2 (HIP~3829 or Wolf~28).  At a distance of 4.3~pc, this is the closest single white dwarf and, after Sirius~B and Procyon~B, the third nearest.  The estimated $\vra$ gives almost perfect agreement between the EDR3 and \textsc{Hipparcos} 
data ($\chi^2_\text{min}=1.7$ for 4~dof). 

Although the presence of a possible companion has been suggested, deep infrared imaging \citepads{2008MNRAS.386L...5B} and searches for binarity from possible proper-motion anomalies by \citetads{2019A&A...623A..72K} seem to exclude any nearby orbiting companion more massive than 
$2~M_\text{Jup}$.   

The strongly deviant spectroscopic value cannot be reconciled with the astrometric data for any reasonable amount of gravitational redshift.  The spectroscopic number is the one generally quoted in catalogue compilations, entering databases such as \textsc{SIMBAD} 
(\citeads{1995A&AS..114..269D}; \citeads{2006AstL...32..759G}), and it merits closer scrutiny.
In tracing the origins of this quantity, one finds it to come from the General Catalogue of Stellar Radial Velocities \citepads{1953GCRV..C......0W}.  Examining the remarks in the printed version of this catalogue,%
\footnote{For example~\url{http://publicationsonline.carnegiescience.edu/publications_online/stellar_radial_velocities.pdf}}
one finds this value to arise from an observation made at Mt.~Wilson Observatory, in a listing of stars with large radial velocities by \citetads{1926PASP...38..121A}.  In that work, data for most 
stars are based upon several spectrograms, although `in a few cases of exceptional interest', even a single observation was included. The stars are typically of visual magnitude 7 to 9, with van Maanen~2 being the faintest of that sample at $V=12.3$ (the second faintest had $V=10.8$).  
Only one exposure was recorded with the lowest resolution camera, and the deduced value is further marked as `subject to especial uncertainty'. Clearly, this cannot be seen as a precise or reliable measurement.

Together with other white dwarfs, van Maanen~2 was later observed with the Hale telescope on Palomar by \citetads{1967ApJ...149..283G}, yielding $\vrs=+54$~km\,s$^{-1}$, a value those authors classified as `reliable'.  In a somewhat later reexamination, \citetads{1972ApJ...173..377G} 
arrived at $+39$~km\,s$^{-1}$ from measuring numerous spectrograms, but with differences between various spectral lines.  For such faint sources with very diffuse spectral lines, measured values may fluctuate greatly depending on the spectrograph dispersion and the depth of photographic exposure \citepads{1954AJ.....59..322G}. Even more precise spectral recordings face challenges: by using only the \ion{Ca}{ii} K line, \citetads{1993AJ....105.1033A} obtained $\vrs=+15$~km\,s$^{-1}$, but a fit to both the H and K lines instead gave $+54$~km\,s$^{-1}$, probably influenced by pressure shifts in the white-dwarf atmosphere.

The large proper motion of van Maanen~2 caused \citetads[][page 287]{1932BAN.....6..249O}, 
expanding on an earlier suggestion by \citetads{1931Natur.127..661R}, to suggest the use of secular 
changes in proper motion to distinguish the gravitational redshift from actual radial motion, envisioned to be feasible after 30 or 40 years of observations.   
Using the long-focus Sproul refractor during 1937--1970, \citetads{1971IAUS...42...32V} derived a value for the astrometric velocity, with a significant uncertainty but remarkably consistent 
with the early spectroscopic measurements.  However, \citetads{1974AJ.....79..815G} pointed out an error in those calculations. Using images from even longer time series with the Allegheny refractor, they deduced a new value with an astrometric velocity of $+6\pm 15$~km\,s$^{-1}$ and argued that this implies a plausible gravitational redshift.  The Sproul plate series were continued until 1976, and they were re-measured by \citetads{1978AJ.....83..197H}, who then found an astrometric value of $+25\pm 18$~km\,s$^{-1}$.  

The actual gravitational redshift depends on the white-dwarf mass.  In a spectrophotometric study, \citetads{2012ApJS..199...29G} obtained the mass 0.68$\pm0.02~M_{\odot}$, implying a likely redshift of the order of 40~km\,s$^{-1}$, thus leaving a contribution to the wavelength shift caused 
by the stellar motion that is reasonably consistent with our astrometric value of $-12\pm 7$~km\,s$^{-1}$. 

The misperception of the anomalously large spectroscopic value for this star seems to have originated 
by the quoting and re-quoting of one particular uncertain observation, whereupon its inclusion 
in general catalogues was chosen as being the value that had the greatest number of citations.
Such a case is by no means unique.  For an example where early 19th century visual observations of 
suggested stellar variability were carried over into modern catalogues, 
see \citetads{1993ApJ...403..385D}.

\subsection{HIP~57367 (LAWD~37)}
\label{ssec:lawd37}

Obtaining accurate gravitational redshifts in white dwarfs remains a challenge. 
Only in the particular nearby system of Sirius~B, observed from Hubble Space Telescope, has a precise 
value of $80.65\pm0.77$~km\,s$^{-1}$ been measured for its rather massive white dwarf 
(\citeads{2005MNRAS.362.1134B}; \citeads{2018MNRAS.481.2361J}).

In Table~\ref{table1} the white dwarf with the most precise determination of astrometric radial velocity is LAWD~37 (HIP~57367 or GJ~440) with $\vra=+28\pm 5$~km\,s$^{-1}$. As van Maanen~2, it fits the combined \textsc{Hipparcos} and \textit{Gaia}~EDR3 astrometry almost perfectly and has no known companion (\citeads{2000AJ....119..906S}; 
\citeads{2019A&A...623A..72K}). It does not have a spectroscopic velocity in \textsc{SIMBAD,} but it is clearly an interesting target for the determination of the gravitational redshift if a spectroscopic $\vrs$ is obtained and an improved astrometric value can be derived using future
releases of \textit{Gaia} data. The significance of such data is enhanced by the prospects of measuring its mass from gravitational microlensing (\citeads{2020A&A...640A..83K}; \citeads{2018MNRAS.478L..29M};
\citeyearads{2020MNRAS.498L...6M}).

Based on \textsc{Hipparcos} and \textit{Gaia} DR2 data, \citetads{2019A&A...623A..72K} identified a
proper-motion anomaly for LAWD~37 at a signal-to-noise ratio  of 4.9, which could suggest the
presence of a massive companion. That result, however, is not supported by the present study.

\subsection{HIP~87937 (Barnard's star)}
\label{ssec:Barnard}

This nearby red dwarf yields the most precise velocity value, with an uncertainty comparable to the star's expected gravitational redshift.  At these accuracy levels, differences between spectroscopic and astrometric values may start to provide unique astrophysical information.  A feature of both Barnard's star and other red dwarfs, is the presence of additional effects that are modulating the stellar spectrum.  Precise Doppler monitoring sets limits on possible exoplanets (\citeads{2013ApJ...764..131C}; 
\citeads{2003A&A...403.1077K}; \citeads{2009A&A...505..859Z}), while variability in chromospheric emission lines suggests a long-term activity cycle (\citeads{2012A&A...541A...9G};
\citeads{2019MNRAS.488.5145T}), accompanied by occasional flaring \citepads{2006PASP..118..227P}. `Chromatic' radial velocities obtained from different parts of the spectrum may segregate signatures from magnetic activity, convection, or other effects \citepads{2020A&A...641A..69B}; however, obtaining a zero point for the different wavelength shifts, corresponding to the radial motion of the stellar centre-of-mass, does require astrometry. 

Analogous considerations apply to other red dwarfs including Proxima Centauri and Kapteyn's star. Nearby red dwarfs are numerous (making up 35 out of the 55 entries in Table~\ref{table1}), and they can be expected to display statistically similar properties.  Even if an adequate accuracy cannot yet be obtained for each of them, there is the prospect of averaging astrometric and spectroscopic velocities for groups of such M-dwarfs to identify features in their spectroscopic signatures once adequate data are available for a sufficient number of them.  As mentioned in connection with Eq.~(\ref{eq07}), the possible presence of perturbing companions to some of the stars should not systematically bias the mean astrometric radial velocity for such a group.

\subsection{HIP~104214 + 104217 (61 Cyg~AB)}
\label{ssec:61cyg}

For the binary system 61~Cyg~AB, we derived the mass ratio $M_\text{B}/M_\text{A}=0.758\pm 0.049$
from the combined \textsc{Hipparcos} and \textit{Gaia} EDR3 astrometry. This gives a significantly better fit to the combined astrometry than the higher mass ratio 0.877 estimated from evolutionary models by \citetads{2008A&A...488..667K}. As discussed therein, the masses of the components are not well constrained by observations and the mass ratio derived here is well within the range of previous dynamical estimates. However, both components are very bright for \textit{Gaia} ($G=4.77$ and $5.45$~mag) and the astrometry in EDR3 is known to degrade rapidly for $G\lesssim 6$~mag \citepads{2018A&A...616A...2L}, which could bias our result.

In \citetads{1999A&A...348.1040D}, we derived an astrometric radial velocity for 61~Cyg of $\vra=-68.0\pm 11.1$~km\,s$^{-1}$. This was obtained from a combination of \textsc{Hipparcos} data and the visual observations by \citetads{1838AN.....16...65B}, made in 1837--38 as part of his successful campaign to measure the parallax of the system. Considering that the perspective effect in position increases quadratically with the temporal baseline, it might still be relevant to consider Bessel's measurements in combination with the \textit{Gaia} and \textsc{Hipparcos} data. However, we have not been able to obtain a consistent solution including all three kinds of data. If this inconsistency would be caused by the problem in EDR3 mentioned above for the very bright stars, the results in Table~\ref{table2} for this system should be viewed with caution.

\section{Conclusions}

The feasibility of determining radial velocities from purely geometric measurements was already realised long ago, along with its ensuing observational demands.  In evaluating the observability of van Maanen's star, \citetads{1931Natur.127..661R} wrote: `in a century or less the true radial velocity of a star of such large proper motions and parallax as this could be found from the second order term in the proper motion'.  Accuracies realised in space astrometry now permit such measurements over more manageable timescales. The most precise results should be obtainable for the nearest stars but, unfortunately, the brighter of those are still difficult to observe.  Proxima Centauri yielded a precise value (Table~\ref{table1}), but the other components of the $\alpha$\,Cen system were too bright to be included in \textit{Gaia} EDR3 and they 
may have to be tied to the \textit{Gaia} reference frame by means of other instruments
(e.g.\ \citeads{2021arXiv210410086A}) or await future exploration of space observations.
Already now, the method permits one to estimate gravitational redshifts of white dwarfs and to set limits on convective shifts in other stars.

\begin{acknowledgements}
This work has made use of data from the European Space Agency (ESA) mission {\it Gaia} (\url{https://www.cosmos.esa.int/gaia}), processed by the {\it Gaia} Data Processing and Analysis Consortium (DPAC, \url{https://www.cosmos.esa.int/web/gaia/dpac/consortium}). Funding for the DPAC has been provided by national institutions, in particular the institutions participating in the {\it Gaia} Multilateral Agreement. The work by LL is supported by the Swedish National Space Agency. The work by DD is supported by grants from The Royal Physiographic Society of Lund. This research has made use of the SIMBAD database, operated at CDS, Strasbourg, France. Diagrams were produced using the astronomy-oriented data handling and visualisation software TOPCAT \citepads{2005ASPC..347...29T}.  Use was made of NASA’s ADS Bibliographic Services and the arXiv$^{\circledR}$ distribution service. We thank the referee, Pierre Kervella, for his detailed and insightful comments.

\end{acknowledgements}

\bibliographystyle{aa} 
\bibliography{refs} 

\begin{thebibliography}{63}
\expandafter\ifx\csname natexlab\endcsname\relax\def\natexlab#1{#1}\fi

\bibitem[{{Aannestad} {et~al.}(1993){Aannestad}, {Kenyon}, {Hammond}, \&
  {Sion}}]{1993AJ....105.1033A}
{Aannestad}, P.~A., {Kenyon}, S.~J., {Hammond}, G.~L., \& {Sion}, E.~M. 1993,
  \aj, 105, 1033

\bibitem[{{Adams} \& {Joy}(1926)}]{1926PASP...38..121A}
{Adams}, W.~S. \& {Joy}, A.~H. 1926, \pasp, 38, 121

\bibitem[{{Akeson} {et~al.}(2021){Akeson}, {Beichman}, {Kervella}, {Fomalont},
  \& {Benedict}}]{2021arXiv210410086A}
{Akeson}, R., {Beichman}, C., {Kervella}, P., {Fomalont}, E., \& {Benedict},
  G.~F. 2021, arXiv e-prints, arXiv:2104.10086

\bibitem[{{Baroch} {et~al.}(2020){Baroch}, {Morales}, {Ribas}, {Herrero},
  {Rosich}, {Perger}, {Anglada-Escud{\'e}}, {Reiners}, {Caballero},
  {Quirrenbach}, {Amado}, {Jeffers}, {Cifuentes}, {Passegger}, {Schweitzer},
  {Lafarga}, {Bauer}, {B{\'e}jar}, {Colom{\'e}}, {Cort{\'e}s-Contreras},
  {Dreizler}, {Galad{\'\i}-Enr{\'\i}quez}, {Hatzes}, {Henning}, {Kaminski},
  {K{\"u}rster}, {Montes}, {Rodr{\'\i}guez-L{\'o}pez}, \&
  {Zechmeister}}]{2020A&A...641A..69B}
{Baroch}, D., {Morales}, J.~C., {Ribas}, I., {et~al.} 2020, \aap, 641, A69

\bibitem[{{Barstow} {et~al.}(2005){Barstow}, {Bond}, {Holberg}, {Burleigh},
  {Hubeny}, \& {Koester}}]{2005MNRAS.362.1134B}
{Barstow}, M.~A., {Bond}, H.~E., {Holberg}, J.~B., {et~al.} 2005, \mnras, 362,
  1134

\bibitem[{{Bessel}(1838)}]{1838AN.....16...65B}
{Bessel}, F.~W. 1838, Astronomische Nachrichten, 16, 65

\bibitem[{{Brandt}(2018)}]{2018ApJS..239...31B}
{Brandt}, T.~D. 2018, \apjs, 239, 31

\bibitem[{{Brandt}(2021)}]{2021arXiv210511662B}
{Brandt}, T.~D. 2021, arXiv e-prints, arXiv:2105.11662

\bibitem[{{Burleigh} {et~al.}(2008){Burleigh}, {Clarke}, {Hogan}, {Brinkworth},
  {Bergeron}, {Dufour}, {Dobbie}, {Levan}, {Hodgkin}, {Hoard}, \&
  {Wachter}}]{2008MNRAS.386L...5B}
{Burleigh}, M.~R., {Clarke}, F.~J., {Hogan}, E., {et~al.} 2008, \mnras, 386, L5

\bibitem[{{Butkevich} \& {Lindegren}(2014)}]{2014A&A...570A..62B}
{Butkevich}, A.~G. \& {Lindegren}, L. 2014, \aap, 570, A62

\bibitem[{{Cantat-Gaudin} \& {Brandt}(2021)}]{2021A&A...649A.124C}
{Cantat-Gaudin}, T. \& {Brandt}, T.~D. 2021, \aap, 649, A124

\bibitem[{{Choi} {et~al.}(2013){Choi}, {McCarthy}, {Marcy}, {Howard},
  {Fischer}, {Johnson}, {Isaacson}, \& {Wright}}]{2013ApJ...764..131C}
{Choi}, J., {McCarthy}, C., {Marcy}, G.~W., {et~al.} 2013, \apj, 764, 131

\bibitem[{{de Bruijne} {et~al.}(2001){de Bruijne}, {Hoogerwerf}, \& {de
  Zeeuw}}]{2001A&A...367..111D}
{de Bruijne}, J.~H.~J., {Hoogerwerf}, R., \& {de Zeeuw}, P.~T. 2001, \aap, 367,
  111

\bibitem[{{Dravins} {et~al.}(1999){Dravins}, {Lindegren}, \&
  {Madsen}}]{1999A&A...348.1040D}
{Dravins}, D., {Lindegren}, L., \& {Madsen}, S. 1999, \aap, 348, 1040

\bibitem[{{Dravins} {et~al.}(1993){Dravins}, {Lindegren}, {Nordlund}, \&
  {Vandenberg}}]{1993ApJ...403..385D}
{Dravins}, D., {Lindegren}, L., {Nordlund}, A., \& {Vandenberg}, D.~A. 1993,
  \apj, 403, 385

\bibitem[{{Duflot} {et~al.}(1995){Duflot}, {Figon}, \&
  {Meyssonnier}}]{1995A&AS..114..269D}
{Duflot}, M., {Figon}, P., \& {Meyssonnier}, N. 1995, \aaps, 114, 269

\bibitem[{{Gaia Collaboration} {et~al.}(2021){Gaia Collaboration}, {Brown},
  {Vallenari}, {Prusti}, {de Bruijne}, {Babusiaux}, {Biermann}, {Creevey},
  {Evans}, {Eyer}, {Hutton}, {Jansen}, {Jordi}, {Klioner}, {Lammers},
  {Lindegren}, {Luri}, {Mignard}, {Panem}, {Pourbaix}, {Randich}, {Sartoretti},
  {Soubiran}, {Walton}, {Arenou}, {Bailer-Jones}, {Bastian}, {Cropper},
  {Drimmel}, {Katz}, {Lattanzi}, {van Leeuwen}, {Bakker}, {Cacciari},
  {Casta{\~n}eda}, {De Angeli}, {Ducourant}, {Fabricius}, {Fouesneau},
  {Fr{\'e}mat}, {Guerra}, {Guerrier}, {Guiraud}, {Jean-Antoine Piccolo},
  {Masana}, {Messineo}, {Mowlavi}, {Nicolas}, {Nienartowicz}, {Pailler},
  {Panuzzo}, {Riclet}, {Roux}, {Seabroke}, {Sordo}, {Tanga}, {Th{\'e}venin},
  {Gracia-Abril}, {Portell}, {Teyssier}, {Altmann}, {Andrae}, {Bellas-Velidis},
  {Benson}, {Berthier}, {Blomme}, {Brugaletta}, {Burgess}, {Busso}, {Carry},
  {Cellino}, {Cheek}, {Clementini}, {Damerdji}, {Davidson}, {Delchambre},
  {Dell'Oro}, {Fern{\'a}ndez-Hern{\'a}ndez}, {Galluccio}, {Garc{\'\i}a-Lario},
  {Garcia-Reinaldos}, {Gonz{\'a}lez-N{\'u}{\~n}ez}, {Gosset}, {Haigron},
  {Halbwachs}, {Hambly}, {Harrison}, {Hatzidimitriou}, {Heiter},
  {Hern{\'a}ndez}, {Hestroffer}, {Hodgkin}, {Holl}, {Jan{\ss}en}, {Jevardat de
  Fombelle}, {Jordan}, {Krone-Martins}, {Lanzafame}, {L{\"o}ffler}, {Lorca},
  {Manteiga}, {Marchal}, {Marrese}, {Moitinho}, {Mora}, {Muinonen}, {Osborne},
  {Pancino}, {Pauwels}, {Petit}, {Recio-Blanco}, {Richards}, {Riello},
  {Rimoldini}, {Robin}, {Roegiers}, {Rybizki}, {Sarro}, {Siopis}, {Smith},
  {Sozzetti}, {Ulla}, {Utrilla}, {van Leeuwen}, {van Reeven}, {Abbas}, {Abreu
  Aramburu}, {Accart}, {Aerts}, {Aguado}, {Ajaj}, {Altavilla}, {{\'A}lvarez},
  {{\'A}lvarez Cid-Fuentes}, {Alves}, {Anderson}, {Anglada Varela}, {Antoja},
  {Audard}, {Baines}, {Baker}, {Balaguer-N{\'u}{\~n}ez}, {Balbinot}, {Balog},
  {Barache}, {Barbato}, {Barros}, {Barstow}, {Bartolom{\'e}}, {Bassilana},
  {Bauchet}, {Baudesson-Stella}, {Becciani}, {Bellazzini}, {Bernet}, {Bertone},
  {Bianchi}, {Blanco-Cuaresma}, {Boch}, {Bombrun}, {Bossini}, {Bouquillon},
  {Bragaglia}, {Bramante}, {Breedt}, {Bressan}, {Brouillet}, {Bucciarelli},
  {Burlacu}, {Busonero}, {Butkevich}, {Buzzi}, {Caffau}, {Cancelliere},
  {C{\'a}novas}, {Cantat-Gaudin}, {Carballo}, {Carlucci}, {Carnerero},
  {Carrasco}, {Casamiquela}, {Castellani}, {Castro-Ginard}, {Castro Sampol},
  {Chaoul}, {Charlot}, {Chemin}, {Chiavassa}, {Cioni}, {Comoretto}, {Cooper},
  {Cornez}, {Cowell}, {Crifo}, {Crosta}, {Crowley}, {Dafonte}, {Dapergolas},
  {David}, {David}, {de Laverny}, {De Luise}, {De March}, {De Ridder}, {de
  Souza}, {de Teodoro}, {de Torres}, {del Peloso}, {del Pozo}, {Delbo},
  {Delgado}, {Delgado}, {Delisle}, {Di Matteo}, {Diakite}, {Diener},
  {Distefano}, {Dolding}, {Eappachen}, {Edvardsson}, {Enke}, {Esquej}, {Fabre},
  {Fabrizio}, {Faigler}, {Fedorets}, {Fernique}, {Fienga}, {Figueras},
  {Fouron}, {Fragkoudi}, {Fraile}, {Franke}, {Gai}, {Garabato},
  {Garcia-Gutierrez}, {Garc{\'\i}a-Torres}, {Garofalo}, {Gavras}, {Gerlach},
  {Geyer}, {Giacobbe}, {Gilmore}, {Girona}, {Giuffrida}, {Gomel}, {Gomez},
  {Gonzalez-Santamaria}, {Gonz{\'a}lez-Vidal}, {Granvik},
  {Guti{\'e}rrez-S{\'a}nchez}, {Guy}, {Hauser}, {Haywood}, {Helmi}, {Hidalgo},
  {Hilger}, {H{\l}adczuk}, {Hobbs}, {Holland}, {Huckle}, {Jasniewicz},
  {Jonker}, {Juaristi Campillo}, {Julbe}, {Karbevska}, {Kervella}, {Khanna},
  {Kochoska}, {Kontizas}, {Kordopatis}, {Korn}, {Kostrzewa-Rutkowska},
  {Kruszy{\'n}ska}, {Lambert}, {Lanza}, {Lasne}, {Le Campion}, {Le Fustec},
  {Lebreton}, {Lebzelter}, {Leccia}, {Leclerc}, {Lecoeur-Taibi}, {Liao},
  {Licata}, {Lindstr{\o}m}, {Lister}, {Livanou}, {Lobel}, {Madrero Pardo},
  {Managau}, {Mann}, {Marchant}, {Marconi}, {Marcos Santos}, {Marinoni},
  {Marocco}, {Marshall}, {Martin Polo}, {Mart{\'\i}n-Fleitas}, {Masip},
  {Massari}, {Mastrobuono-Battisti}, {Mazeh}, {McMillan}, {Messina},
  {Michalik}, {Millar}, {Mints}, {Molina}, {Molinaro}, {Moln{\'a}r},
  {Montegriffo}, {Mor}, {Morbidelli}, {Morel}, {Morris}, {Mulone}, {Munoz},
  {Muraveva}, {Murphy}, {Musella}, {Noval}, {Ord{\'e}novic}, {Orr{\`u}},
  {Osinde}, {Pagani}, {Pagano}, {Palaversa}, {Palicio}, {Panahi}, {Pawlak},
  {Pe{\~n}alosa Esteller}, {Penttil{\"a}}, {Piersimoni}, {Pineau}, {Plachy},
  {Plum}, {Poggio}, {Poretti}, {Poujoulet}, {Pr{\v{s}}a}, {Pulone}, {Racero},
  {Ragaini}, {Rainer}, {Raiteri}, {Rambaux}, {Ramos}, {Ramos-Lerate}, {Re
  Fiorentin}, {Regibo}, {Reyl{\'e}}, {Ripepi}, {Riva}, {Rixon}, {Robichon},
  {Robin}, {Roelens}, {Rohrbasser}, {Romero-G{\'o}mez}, {Rowell}, {Royer},
  {Rybicki}, {Sadowski}, {Sagrist{\`a} Sell{\'e}s}, {Sahlmann}, {Salgado},
  {Salguero}, {Samaras}, {Sanchez Gimenez}, {Sanna}, {Santove{\~n}a},
  {Sarasso}, {Schultheis}, {Sciacca}, {Segol}, {Segovia}, {S{\'e}gransan},
  {Semeux}, {Shahaf}, {Siddiqui}, {Siebert}, {Siltala}, {Slezak}, {Smart},
  {Solano}, {Solitro}, {Souami}, {Souchay}, {Spagna}, {Spoto}, {Steele},
  {Steidelm{\"u}ller}, {Stephenson}, {S{\"u}veges}, {Szabados}, {Szegedi-Elek},
  {Taris}, {Tauran}, {Taylor}, {Teixeira}, {Thuillot}, {Tonello}, {Torra},
  {Torra}, {Turon}, {Unger}, {Vaillant}, {van Dillen}, {Vanel}, {Vecchiato},
  {Viala}, {Vicente}, {Voutsinas}, {Weiler}, {Wevers}, {Wyrzykowski}, {Yoldas},
  {Yvard}, {Zhao}, {Zorec}, {Zucker}, {Zurbach}, \&
  {Zwitter}}]{2021A&A...649A...1G}
{Gaia Collaboration}, {Brown}, A.~G.~A., {Vallenari}, A., {et~al.} 2021, \aap,
  649, A1

\bibitem[{{Gatewood} \& {Russell}(1974)}]{1974AJ.....79..815G}
{Gatewood}, G. \& {Russell}, J. 1974, \aj, 79, 815

\bibitem[{{Giammichele} {et~al.}(2012){Giammichele}, {Bergeron}, \&
  {Dufour}}]{2012ApJS..199...29G}
{Giammichele}, N., {Bergeron}, P., \& {Dufour}, P. 2012, \apjs, 199, 29

\bibitem[{{Gomes da Silva} {et~al.}(2012){Gomes da Silva}, {Santos}, {Bonfils},
  {Delfosse}, {Forveille}, {Udry}, {Dumusque}, \&
  {Lovis}}]{2012A&A...541A...9G}
{Gomes da Silva}, J., {Santos}, N.~C., {Bonfils}, X., {et~al.} 2012, \aap, 541,
  A9

\bibitem[{{Gontcharov}(2006)}]{2006AstL...32..759G}
{Gontcharov}, G.~A. 2006, Astronomy Letters, 32, 759

\bibitem[{{Greenstein}(1954)}]{1954AJ.....59..322G}
{Greenstein}, J.~L. 1954, \aj, 59, 322

\bibitem[{{Greenstein}(1972)}]{1972ApJ...173..377G}
{Greenstein}, J.~L. 1972, \apj, 173, 377

\bibitem[{{Greenstein} \& {Trimble}(1967)}]{1967ApJ...149..283G}
{Greenstein}, J.~L. \& {Trimble}, V.~L. 1967, \apj, 149, 283

\bibitem[{{Hershey}(1978)}]{1978AJ.....83..197H}
{Hershey}, J.~L. 1978, \aj, 83, 197

\bibitem[{{Joyce} {et~al.}(2018){Joyce}, {Barstow}, {Holberg}, {Bond},
  {Casewell}, \& {Burleigh}}]{2018MNRAS.481.2361J}
{Joyce}, S.~R.~G., {Barstow}, M.~A., {Holberg}, J.~B., {et~al.} 2018, \mnras,
  481, 2361

\bibitem[{{Kervella} {et~al.}(2019){Kervella}, {Arenou}, {Mignard}, \&
  {Th{\'e}venin}}]{2019A&A...623A..72K}
{Kervella}, P., {Arenou}, F., {Mignard}, F., \& {Th{\'e}venin}, F. 2019, \aap,
  623, A72

\bibitem[{{Kervella} {et~al.}(2008){Kervella}, {M{\'e}rand}, {Pichon},
  {Th{\'e}venin}, {Heiter}, {Bigot}, {ten Brummelaar}, {McAlister}, {Ridgway},
  {Turner}, {Sturmann}, {Sturmann}, {Goldfinger}, \&
  {Farrington}}]{2008A&A...488..667K}
{Kervella}, P., {M{\'e}rand}, A., {Pichon}, B., {et~al.} 2008, \aap, 488, 667

\bibitem[{{Kl{\"u}ter} {et~al.}(2020){Kl{\"u}ter}, {Bastian}, \&
  {Wambsganss}}]{2020A&A...640A..83K}
{Kl{\"u}ter}, J., {Bastian}, U., \& {Wambsganss}, J. 2020, \aap, 640, A83

\bibitem[{{Kovalevsky} {et~al.}(1997){Kovalevsky}, {Lindegren}, {Perryman},
  {Hemenway}, {Johnston}, {Kislyuk}, {Lestrade}, {Morrison}, {Platais}, \&
  {R{\"o}ser}}]{1997A&A...323..620K}
{Kovalevsky}, J., {Lindegren}, L., {Perryman}, M.~A.~C., {et~al.} 1997, \aap,
  323, 620

\bibitem[{{K{\"u}rster} {et~al.}(2003){K{\"u}rster}, {Endl}, {Rouesnel}, {Els},
  {Kaufer}, {Brillant}, {Hatzes}, {Saar}, \& {Cochran}}]{2003A&A...403.1077K}
{K{\"u}rster}, M., {Endl}, M., {Rouesnel}, F., {et~al.} 2003, \aap, 403, 1077

\bibitem[{{Le{\~a}o} {et~al.}(2019){Le{\~a}o}, {Pasquini}, {Ludwig}, \& {de
  Medeiros}}]{2019MNRAS.483.5026L}
{Le{\~a}o}, I.~C., {Pasquini}, L., {Ludwig}, H.~G., \& {de Medeiros}, J.~R.
  2019, \mnras, 483, 5026

\bibitem[{{Lindegren} {et~al.}(2021){Lindegren}, {Bastian}, {Biermann},
  {Bombrun}, {de Torres}, {Gerlach}, {Geyer}, {Hern{\'a}ndez}, {Hilger},
  {Hobbs}, {Klioner}, {Lammers}, {McMillan}, {Ramos-Lerate},
  {Steidelm{\"u}ller}, {Stephenson}, \& {van Leeuwen}}]{2021A&A...649A...4L}
{Lindegren}, L., {Bastian}, U., {Biermann}, M., {et~al.} 2021, \aap, 649, A4

\bibitem[{{Lindegren} \& {Dravins}(2003)}]{2003A&A...401.1185L}
{Lindegren}, L. \& {Dravins}, D. 2003, \aap, 401, 1185

\bibitem[{{Lindegren} {et~al.}(2018){Lindegren}, {Hern{\'a}ndez}, {Bombrun},
  {Klioner}, {Bastian}, {Ramos-Lerate}, {de Torres}, {Steidelm{\"u}ller},
  {Stephenson}, {Hobbs}, {Lammers}, {Biermann}, {Geyer}, {Hilger}, {Michalik},
  {Stampa}, {McMillan}, {Casta{\~n}eda}, {Clotet}, {Comoretto}, {Davidson},
  {Fabricius}, {Gracia}, {Hambly}, {Hutton}, {Mora}, {Portell}, {van Leeuwen},
  {Abbas}, {Abreu}, {Altmann}, {Andrei}, {Anglada}, {Balaguer-N{\'u}{\~n}ez},
  {Barache}, {Becciani}, {Bertone}, {Bianchi}, {Bouquillon}, {Bourda},
  {Br{\"u}semeister}, {Bucciarelli}, {Busonero}, {Buzzi}, {Cancelliere},
  {Carlucci}, {Charlot}, {Cheek}, {Crosta}, {Crowley}, {de Bruijne}, {de
  Felice}, {Drimmel}, {Esquej}, {Fienga}, {Fraile}, {Gai}, {Garralda},
  {Gonz{\'a}lez-Vidal}, {Guerra}, {Hauser}, {Hofmann}, {Holl}, {Jordan},
  {Lattanzi}, {Lenhardt}, {Liao}, {Licata}, {Lister}, {L{\"o}ffler},
  {Marchant}, {Martin-Fleitas}, {Messineo}, {Mignard}, {Morbidelli}, {Poggio},
  {Riva}, {Rowell}, {Salguero}, {Sarasso}, {Sciacca}, {Siddiqui}, {Smart},
  {Spagna}, {Steele}, {Taris}, {Torra}, {van Elteren}, {van Reeven}, \&
  {Vecchiato}}]{2018A&A...616A...2L}
{Lindegren}, L., {Hern{\'a}ndez}, J., {Bombrun}, A., {et~al.} 2018, \aap, 616,
  A2

\bibitem[{{Lindegren} {et~al.}(2000){Lindegren}, {Madsen}, \&
  {Dravins}}]{2000A&A...356.1119L}
{Lindegren}, L., {Madsen}, S., \& {Dravins}, D. 2000, \aap, 356, 1119

\bibitem[{{Madsen} {et~al.}(2002){Madsen}, {Dravins}, \&
  {Lindegren}}]{2002A&A...381..446M}
{Madsen}, S., {Dravins}, D., \& {Lindegren}, L. 2002, \aap, 381, 446

\bibitem[{{Madsen} {et~al.}(2003){Madsen}, {Dravins}, {Ludwig}, \&
  {Lindegren}}]{2003A&A...411..581M}
{Madsen}, S., {Dravins}, D., {Ludwig}, H.-G., \& {Lindegren}, L. 2003, \aap,
  411, 581

\bibitem[{{Mann} {et~al.}(2015){Mann}, {Feiden}, {Gaidos}, {Boyajian}, \& {von
  Braun}}]{2015ApJ...804...64M}
{Mann}, A.~W., {Feiden}, G.~A., {Gaidos}, E., {Boyajian}, T., \& {von Braun},
  K. 2015, \apj, 804, 64

\bibitem[{{McGill} {et~al.}(2020){McGill}, {Everall}, {Boubert}, \&
  {Smith}}]{2020MNRAS.498L...6M}
{McGill}, P., {Everall}, A., {Boubert}, D., \& {Smith}, L.~C. 2020, \mnras,
  498, L6

\bibitem[{{McGill} {et~al.}(2018){McGill}, {Smith}, {Evans}, {Belokurov}, \&
  {Smart}}]{2018MNRAS.478L..29M}
{McGill}, P., {Smith}, L.~C., {Evans}, N.~W., {Belokurov}, V., \& {Smart},
  R.~L. 2018, \mnras, 478, L29

\bibitem[{{Meunier}(2021)}]{2021arXiv210406072M}
{Meunier}, N. 2021, arXiv e-prints, arXiv:2104.06072

\bibitem[{{Michalik} {et~al.}(2014){Michalik}, {Lindegren}, {Hobbs}, \&
  {Lammers}}]{2014A&A...571A..85M}
{Michalik}, D., {Lindegren}, L., {Hobbs}, D., \& {Lammers}, U. 2014, \aap, 571,
  A85

\bibitem[{{Moschella} {et~al.}(2021){Moschella}, {Slone}, {Dror}, {Cantiello},
  \& {Perets}}]{2021arXiv210201079M}
{Moschella}, M., {Slone}, O., {Dror}, J.~A., {Cantiello}, M., \& {Perets},
  H.~B. 2021, arXiv e-prints, arXiv:2102.01079

\bibitem[{{Oort}(1932)}]{1932BAN.....6..249O}
{Oort}, J.~H. 1932, \bain, 6, 249

\bibitem[{{Pasquini} {et~al.}(2011){Pasquini}, {Melo}, {Chavero}, {Dravins},
  {Ludwig}, {Bonifacio}, \& {de La Reza}}]{2011A&A...526A.127P}
{Pasquini}, L., {Melo}, C., {Chavero}, C., {et~al.} 2011, \aap, 526, A127

\bibitem[{{Paulson} {et~al.}(2006){Paulson}, {Allred}, {Anderson}, {Hawley},
  {Cochran}, \& {Yelda}}]{2006PASP..118..227P}
{Paulson}, D.~B., {Allred}, J.~C., {Anderson}, R.~B., {et~al.} 2006, \pasp,
  118, 227

\bibitem[{{Perryman} {et~al.}(1997){Perryman}, {Lindegren}, {Kovalevsky},
  {Hog}, {Bastian}, {Bernacca}, {Creze}, {Donati}, {Grenon}, {Grewing}, {van
  Leeuwen}, {van der Marel}, {Mignard}, {Murray}, {Le Poole}, {Schrijver},
  {Turon}, {Arenou}, {Froeschle}, \& {Petersen}}]{1997A&A...323L..49P}
{Perryman}, M.~A.~C., {Lindegren}, L., {Kovalevsky}, J., {et~al.} 1997, \aap,
  500, 501

\bibitem[{{Pourbaix} {et~al.}(2002){Pourbaix}, {Nidever}, {McCarthy}, {Butler},
  {Tinney}, {Marcy}, {Jones}, {Penny}, {Carter}, {Bouchy}, {Pepe}, {Hearnshaw},
  {Skuljan}, {Ramm}, \& {Kent}}]{2002A&A...386..280P}
{Pourbaix}, D., {Nidever}, D., {McCarthy}, C., {et~al.} 2002, \aap, 386, 280

\bibitem[{Press {et~al.}(2007)Press, Teukolsky, Vetterling, \&
  Flannery}]{book:nr3}
Press, W., Teukolsky, S., Vetterling, W., \& Flannery, B. 2007, Numerical
  Recipes: The Art of Scientific Computing, 3rd edn. (Cambridge University
  Press)

\bibitem[{{Ristenpart}(1902)}]{1902Vierteljahrsschrift....37...242S}
{Ristenpart}, F. 1902, Vierteljahrsschrift der Astron.\ Ges., 37, 242

\bibitem[{{Russell} \& {Atkinson}(1931)}]{1931Natur.127..661R}
{Russell}, H.~N. \& {Atkinson}, R.~D. 1931, \nat, 127, 661

\bibitem[{{Schlesinger}(1917)}]{1917AJ.....30..137S}
{Schlesinger}, F. 1917, \aj, 30, 137

\bibitem[{{Schroeder} {et~al.}(2000){Schroeder}, {Golimowski}, {Brukardt},
  {Burrows}, {Caldwell}, {Fastie}, {Ford}, {Hesman}, {Kletskin}, {Krist},
  {Royle}, \& {Zubrowski}}]{2000AJ....119..906S}
{Schroeder}, D.~J., {Golimowski}, D.~A., {Brukardt}, R.~A., {et~al.} 2000, \aj,
  119, 906

\bibitem[{{Seeliger}(1900)}]{1900AN....154...65S}
{Seeliger}, H. 1900, Astronomische Nachrichten, 154, 65

\bibitem[{{Taylor}(2005)}]{2005ASPC..347...29T}
{Taylor}, M.~B. 2005, in Astronomical Society of the Pacific Conference Series,
  Vol. 347, Astronomical Data Analysis Software and Systems XIV, ed.
  P.~{Shopbell}, M.~{Britton}, \& R.~{Ebert}, 29

\bibitem[{{Toledo-Padr{\'o}n} {et~al.}(2019){Toledo-Padr{\'o}n}, {Gonz{\'a}lez
  Hern{\'a}ndez}, {Rodr{\'\i}guez-L{\'o}pez}, {Su{\'a}rez Mascare{\~n}o},
  {Rebolo}, {Butler}, {Ribas}, {Anglada-Escud{\'e}}, {Johnson}, {Reiners},
  {Caballero}, {Quirrenbach}, {Amado}, {B{\'e}jar}, {Morales}, {Perger},
  {Jeffers}, {Vogt}, {Teske}, {Shectman}, {Crane}, {D{\'\i}az}, {Arriagada},
  {Holden}, {Burt}, {Rodr{\'\i}guez}, {Herrero}, {Murgas}, {Pall{\'e}},
  {Morales}, {L{\'o}pez-Gonz{\'a}lez}, {D{\'\i}ez Alonso}, {Tuomi}, {Kiraga},
  {Engle}, {Guinan}, {Strachan}, {Aceituno}, {Aceituno}, {Casanova},
  {Mart{\'\i}n-Ruiz}, {Montes}, {Ortiz}, {Sota}, {Briol}, {Barbieri},
  {Cervini}, {Deldem}, {Dubois}, {Hambsch}, {Harris}, {Kotnik}, {Logie},
  {Lopez}, {McNeely}, {Ogmen}, {P{\'e}rez}, {Rau}, {Rodr{\'\i}guez}, {Urquijo},
  \& {Vanaverbeke}}]{2019MNRAS.488.5145T}
{Toledo-Padr{\'o}n}, B., {Gonz{\'a}lez Hern{\'a}ndez}, J.~I.,
  {Rodr{\'\i}guez-L{\'o}pez}, C., {et~al.} 2019, \mnras, 488, 5145

\bibitem[{{van de Kamp}(1971)}]{1971IAUS...42...32V}
{van de Kamp}, P. 1971, in White Dwarfs, IAU Symp., ed. W.~J. {Luyten},
  Vol.~42, 32

\bibitem[{{van de Kamp}(1977)}]{1977VA.....21..289V}
{van de Kamp}, P. 1977, Vistas in Astronomy, 21, 289

\bibitem[{{van Leeuwen}(2007)}]{2007ASSL..350.....V}
{van Leeuwen}, F. 2007, {Hipparcos, the New Reduction of the Raw Data, AASL},
  Vol. 350 (Springer)

\bibitem[{{Wenger} {et~al.}(2000){Wenger}, {Ochsenbein}, {Egret}, {Dubois},
  {Bonnarel}, {Borde}, {Genova}, {Jasniewicz}, {Lalo{\"e}}, {Lesteven}, \&
  {Monier}}]{2000A&AS..143....9W}
{Wenger}, M., {Ochsenbein}, F., {Egret}, D., {et~al.} 2000, \aaps, 143, 9

\bibitem[{{Wilson}(1953)}]{1953GCRV..C......0W}
{Wilson}, R.~E. 1953, General Catalogue of Stellar Radial Velocities, Carnegie
  Institution of Washington Publ.\ 601, Washington D.C., 344 pp.

\bibitem[{{Zechmeister} {et~al.}(2009){Zechmeister}, {K{\"u}rster}, \&
  {Endl}}]{2009A&A...505..859Z}
{Zechmeister}, M., {K{\"u}rster}, M., \& {Endl}, M. 2009, \aap, 505, 859

\end{thebibliography}

\end{document}